\documentclass[onecolumn,12pt,journal,draftclsnofoot]{IEEEtran}
\usepackage{epsf,psfrag,amssymb,amsfonts,amsmath,graphicx,cite, times}
\usepackage[mathscr]{eucal}

\def\boxit#1{\vbox{\hrule\hbox{\vrule\kern3pt
        \vbox{\kern3pt#1\kern3pt}\kern3pt\vrule}\hrule}}

\def\reals{ { {\rm  I \kern-0.15em R }  } }
\def\complex{ {\,{{\rm C} \kern-0.50em \raise0.20ex {  |}}\, }}

\def\Rbf{{\bf R}}
\def\Sbf{{\bf S}}

\def\Cc{{\cal C}}

\def\Hc{{\cal H}}

\def\Tc{{\cal T}}

\def\defeq{{\stackrel{\Delta}{=}}}
\def\Rxx{\Rbf_{\ssstyle X\kern-.1em X}}

\let\ssstyle=\scriptscriptstyle

\def\ie{{\it i.e.,\ \/}}
\def\Kout{\setbox1=\hbox{\Huge\bf K}\hbox to
1.05\wd1{\hspace{.05\wd1}
}}
\def\Sout{\setbox1=\hbox{\Huge\bf S}\hbox to 1.05\wd1{\hspace{.05\wd1}
}}

\def\ie{{\it i.e.,\ \/}}
\def\defeq{{\stackrel{\Delta}{=}}}

\def\scalefig#1{\epsfxsize #1\textwidth}
\def\nn{{\nonumber}}
\newcommand{\mbbE}{\mathbb{E}}
\newtheorem{lemma}{Lemma}
\newtheorem{theorem}{Theorem}

\begin{document}
\title{\bf \Large Structure and Optimality of the Myopic Policy in\\[0.1em]
Opportunistic Access with Noisy Observations\thanks{This work was supported by the
Army Research Laboratory CTA on Communication and Networks under
Grant DAAD19-01-2-0011 and by the National Science Foundation under
Grants CNS-0627090, ECS-0622200, and CNS-0347621. Part of this work
was presented at {\em the 2nd International Conference on Cognitive Radio Oriented Wireless Networks and Communications (CrownCom)}, August, 2007.}}
\author{Qing Zhao$^*$, ~~~Bhaskar Krishnamachari\thanks{Q. Zhao is with the
Department of Electrical and Computer Engineering, University of
California, Davis, CA 95616. Email:
qzhao@ece.ucdavis.edu. B. Krishnamachari is with the Ming Hsieh Department of Electrical Engineering,
University of Southern California, Los Angeles, CA 90089. Email: bkrishna@usc.edu.}\thanks{$*$ Corresponding author. Phone: 1-530-752-7390. Fax:
1-530-752-8428.}}

\markboth{Submitted to {\em IEEE Transactions on Automatic Control} in February, 2008, revised in August, 2008.}{Zhao and Krishnamachari}

\maketitle%
\thispagestyle{empty}

\vspace{-2em}

\begin{abstract}
A restless multi-armed bandit problem that arises in multichannel
opportunistic communications is considered, where channels are modeled as independent and
identical Gilbert-Elliot channels and channel state observations are subject to errors.
A simple structure of the myopic policy
is established under a certain condition on the false alarm probability of the channel state
detector. It is shown that the myopic policy has a semi-universal structure that reduces
channel selection to a simple
round-robin procedure and obviates the need to know the underlying Markov transition probabilities.
The optimality
of the myopic policy is proved for the case of two channels and conjectured for the general case
based on numerical examples.

\vspace{1em}

\end{abstract}

%\newpage

%\vspace{-0.6em}
\section{Introduction}
\vspace{-0.5em}

\begin{figure}[htb]
\centerline{
\begin{psfrags}
\scalefig{0.9} \psfrag{A}[c]{ $0$} \psfrag{B}[c]{ $1$}
\psfrag{A1}[c]{ (bad)} \psfrag{B1}[c]{ (good)} \psfrag{a}[c]{
$p_{01}$} \psfrag{b}[l]{ $p_{11}$} \psfrag{a1}[r]{
$p_{00}$} \psfrag{b1}[c]{ $p_{10}$}
\scalefig{0.35}\epsfbox{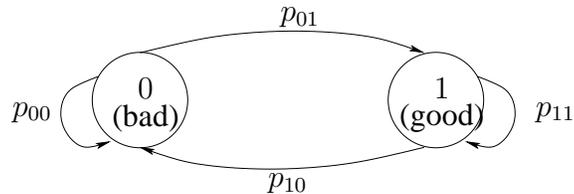}
\end{psfrags}}
\caption{The Gilbert-Elliot channel model.} \label{fig:MC}
\end{figure}

We consider the following stochastic control problem that arises in multichannel
opportunistic communications. Assume that there are $N$ independent
and stochastically identical Gilbert-Elliot channels~\cite{Gilbert:60}. As illustrated in Fig.~\ref{fig:MC},
the state of a channel --- ``good'' or ``bad'' ---
indicates the desirability of accessing this channel and determines the resulting reward.
The transitions between these two states follow a discrete-time Markov chain with transition probabilities $\{p_{ij}\}_{i,j=0,1}$.
This channel model has been commonly used to abstract physical channels with memory
(see \cite{Zorzi&etal:98COM,Johnston&Krishnamurthy:06TWC} and references therein). Consider, for example,
the emerging application of cognitive radios for opportunistic spectrum access where secondary users search in the spectrum for
idle channels temporarily unused by primary users~\cite{Zhao&Sadler:07SPM}.
For this application, the good state represents
an idle channel while the bad state an occupied channel\footnote{When the primary network employs load balancing across channels,
the occupancy processes of all channels can be considered
stochastically identical.}.

In each time slot, a user chooses one of the $N$ channels to sense and subsequently access
if the chosen channel is sensed to be in the good state. Sensing is subject to errors:
a good channel may be sensed as bad and {\it vice versa}. Accessing a good channel
results in a unit reward, and no access or accessing a bad channel leads to zero reward.
The design objective is the optimal sensing policy for channel selection in order to maximize
the expected long-term reward.
This problem can be formulated as a partially observable Markov decision process
(POMDP) for generally correlated channels, or a
restless multi-armed bandit process for
independent channels.

It has been shown in~\cite{tsitsiklis} that obtaining the optimal policy for a general restless
multi-armed bandit problem is PSPACE-hard. For special classes of restless bandit processes,
however, simple structural policies may exist that achieve optimality with low complexity.
As shown in this paper, for the multichannel opportunistic access problem stated
above, the myopic policy for this problem has a simple
and robust structure that reduces channel selection to a simple round-robin procedure when the
false alarm probability of the channel state detector
is below a certain value.
This structure reveals that the myopic policy does not require the knowledge of
the transition probabilities of the Markovian model except the order of $p_{11}$ and $p_{01}$.
The myopic policy thus
automatically tracks variations in the channel model provided that the order of
$p_{11}$ and $p_{01}$ remains unchanged.
Furthermore, exploiting this simple structure, we prove that the myopic policy is optimal for $N=2$.
Numerical examples\footnote{Actions given by the myopic
policy and the optimal policy are compared numerically for randomly chosen $p_{11}$ and $p_{01}$
and $N=3,~4,$ and $5$. All examples show the equivalence between the myopic policy and the optimal
policy.} suggest its optimality for general $N$.

This technical note extends our earlier work in \cite{Zhao&Krish:07CogNet} that assumes
perfect observation of channel states. As shown in Sections~\ref{sec:problem} and
\ref{sec:results}, communication constraints, namely, synchronization in channel selection between
the transmitter and its receiver, require
changes in the problem formulation when observations are imperfect, and uncertainties in the state of sensed channels
complicate the proofs for the structure and optimality of the myopic policy.

\vspace{-0.6em}
\section{Problem Formulation}
\label{sec:problem}

\subsection{System Model}

Let $\Sbf(t)\,\defeq\,[S_1(t), \ldots, S_N(t)]$
denote the channel states, where
$S_n(t) \in \{0\mbox{ (bad), }1\mbox{ (good)}\}$ is the state
of channel $n$ in slot $t$.
At the beginning of each slot, the user first decides which of the $N$ channels to choose
for potential access. Once a channel (say channel $n$) is chosen, the user
detects the channel state, which can be considered as a binary hypothesis test\footnote{We consider
here the nontrivial cases with $p_{01}$ and $p_{11}$ in the open interval of $(0,1)$. When they take
the special value of $0$ or $1$, channel state detection can be simplified. Extensions to such special
cases are straightforward.}:
\begin{equation}
\Hc_0 : S_n(t) = 1 \mbox{~(good)}~~~
\mbox{vs.}~~~ \Hc_1 : S_n(t) = 0 \mbox{~(bad)}.\nn
\end{equation}
The performance of channel state detection is characterized by the
probability of false alarm $\epsilon$ and the probability of miss detection $\delta$:
\begin{equation}
\epsilon \,\defeq\, \Pr\{\mbox{decide }\Hc_1\,|\,\Hc_0\mbox{ is
true}\},~~~
\delta \,\defeq\, \Pr\{\mbox{decide }\Hc_0\,|\,\Hc_1\mbox{ is
true}\}.\nn
\end{equation}
For example, in the application of cognitive radios for opportunistic spectrum access, the user can employ
an energy detector to detect the presence of primary signals. If the measured energy is above a certain
threshold, the channel is detected as bad (\ie busy). Otherwise, the channel is considered idle
and suitable for transmission.
%Clearly, the probability of false alarm can be reduced by lowering the
%detection threshold, but at the price of increasing the probability of miss detection.

The user transmits over the chosen channel if and only if the channel is detected as
in the good state. Thus, one of the following four possible events can occur in each slot:
(i) the chosen channel is good and is correctly detected as such, resulting in a successful transmission;
(ii) a false alarm occurs, and a communication opportunity
is missed; (iii) the chosen channel is bad and is correctly detected; the transmitter refrains from transmitting;
(iv) a miss detection occurs, resulting in a failed transmission.
Only in the first event, a unit reward is accrued in this slot.
The objective is to maximize the average reward
(throughput) over a horizon of $T$ slots by choosing judiciously a
sensing policy that governs channel selection in each slot\footnote{Note that often the design should be subject to
a constraint on the probability of accessing a bad channel, which may cause interference or waste energy. For
example, in the application
of cognitive radios for opportunistic spectrum access, transmitting over a bad (busy) channel leads to a collision
with primary users and should be limited below a prescribed level.
This constrained stochastic control problem requires the joint design of the channel state detector (\ie how to choose
the detection threshold to trade off false alarms with miss detections), the access policy that decides the transmission
probability based on imperfect detection outcome, and
the sensing policy for channel selection. It has been shown in \cite{Chen&etal:08IT} under a general correlated channel model
that the optimal detector is the Neyman-Pearson detector with the probability of miss detection
given by the maximum allowable probability of collision, and the optimal access policy is to simply trust
the detection outcome: transmit if and only if the channel is detected as good.
The optimal sensing policy can then be designed using this optimal detector and the optimal
access policy without the constraint on accessing a bad channel. This is the problem addressed in this paper.}.

Since failed transmissions may occur, acknowledgements are necessary to ensure guaranteed delivery. Specifically,
when the receiver successfully receives a packet (event (i)), it sends an acknowledgement to the transmitter at the end of the slot.
Otherwise, the receiver does nothing, \ie a NAK is defined as the absence of an ACK, which occurs when the transmitter
did not transmit (events (ii) and (iii)) or transmitted over a bad channel (event (iv)).
We assume that acknowledgements are received without error since acknowledgements
are always transmitted over a good/idle channel.

\vspace{-0.6em}
\subsection{Value Function and Belief Update}
\vspace{-0.5em}

While the full system state $\Sbf(t)=[S_1(t),\cdots,S_N(t)]$
is not observable, the user can infer the state from its decision and
observation history. A sufficient statistic for optimal decision making
is given by the conditional probability that each channel
is in state $1$ given all past decisions and observations \cite{Smallwood&Sondik:71OR}.
Referred to as the belief vector (or information state), this sufficient statistic is denoted by
$\Omega(t) \,\defeq\, [\omega_1(t),\cdots,\omega_N(t)]$, where $\omega_i(t)$
is the conditional probability that $S_i(t)=1$. In order to ensure that the user and its
intended receiver tune to the same channel in each slot, channel selections should be
based on common observations: the acknowledgement $K(t)\in\{0\mbox{ (NAK)},~1\mbox{ (ACK)}\}$
in each slot rather than the detection outcome at the transmitter. Given the action $a$ and observation $K_a(t)=k~(k=0,1)$, the belief vector
in slot $t+1$ can be obtained via the Bayes rule.
\begin{equation}
\omega_i(t+1)=\left\{\begin{array}{ll}
p_{11}, &  a=i, K_{a}(t)=1\\
\Gamma(\frac{\epsilon\omega_i(t)}{\epsilon\omega_i(t)+(1-\omega_i(t))}), &  a=i, K_{a}(t)=0\\
\Gamma(\omega_i(t)), & a\neq i\\
\end{array}
\right.,
\label{eq:omega}
\end{equation}
where the operator $\Gamma(\cdot)$ is defined as $\Gamma(x)\defeq xp_{11} + (1-x)p_{01}$.

A sensing policy $\pi$ specifies a sequence of functions $\pi =
[\pi_1, \pi_2, \cdots, \pi_T]$ where $\pi_t$ maps a belief vector
$\Omega(t)$ to a sensing action $a(t)\in\{1,\cdots,N\}$ for slot
$t$. We thus arrive at
the following stochastic control problem.
\begin{equation}
\pi^*=\arg\max_\pi \mbbE_{\pi}\left[\sum_{t=1}^T R_{\pi_t(\Omega(t))}(t)|\Omega(1)\right],
\label{eq:pi*}
\end{equation}
where $R_{\pi_t(\Omega(t))}(t)$ is the reward obtained when the belief is $\Omega(t)$ and
channel $a=\pi_t(\Omega(t))$ is selected, and $\Omega(1)$ is the initial belief vector.
This problem falls into
the general model of POMDP. It can also be considered as a restless
multi-armed bandit problem by treating the belief value of each channel
as the state of each arm of a bandit.

Let $V_t(\Omega)$ be the value function, which represents the maximum expected remaining reward that can be
accrued starting from slot $t$ when the current belief vector is $\Omega$.
We have the following optimality equation.
\begin{eqnarray}
V_T(\Omega)&=&\max_{a=1,\cdots,N} \omega_a (1-\epsilon),\nn\\
V_t(\Omega)&=&\max_{a=1,\cdots,N}\{\omega_a (1-\epsilon)+\omega_a (1-\epsilon)V_{t+1}(\Tc(\Omega|a,1))
+(1-\omega_a (1-\epsilon))V_{t+1}(\Tc(\Omega|a,0))\},\nn
%\label{eq:Vn}
\end{eqnarray}
where $\Tc(\Omega|a,i)$ denotes the updated belief vector for slot $t+1$ after incorporating
action $a$ and observation $K(t)=i$ as given in \eqref{eq:omega}.

In theory, the optimal policy $\pi^*$ can be obtained by solving the above dynamic
program. Unfortunately, this approach is computationally prohibitive
due to the impact of the current action on the future reward and the
uncountable space of the belief vector $\Omega$.

\vspace{-0.6em}
\section{Structure and Optimality of Myopic Policy}
\label{sec:results}
\vspace{-0.5em}

A myopic policy ignores the impact of the current action on the future reward,
focusing solely on maximizing the expected immediate reward $\mbbE[R_a(t)]=\omega_a(t) (1-\epsilon)$.
It is an index policy and is stationary: the mapping from
belief vectors to actions does not change with time $t$.  The myopic action $\hat{a}(t)$ in slot $t$
under belief state $\Omega(t)$ is simply given by
\begin{equation}
\hat{a}(t)=\arg\max_{a=1,\cdots,N} \omega_a(t).
\label{eq:a*}
\end{equation}

In general, obtaining the myopic action in each slot requires
the recursive update of the belief vector $\Omega(t)$ as given in
\eqref{eq:omega}, which requires the knowledge of the transition
probabilities $\{p_{ij}\}$. As shown in Theorem~\ref{thm:structure},
for the problem at hand, the myopic
policy has a simple structure that does not need the update of the
belief vector or the knowledge of the transition
probabilities.

The basic element in the structure of the myopic policy is a circular ordering $\Cc$ of the channels.
For a circular order, the starting point is irrelevant: a circular order
$\Cc=(n_1,n_2,\cdots,n_N)$ is equivalent to
$(n_i,n_{i+1},\cdots, n_N, n_1, n_2,\cdots, n_{i-1})$ for any $1\le i\le N$.

We now introduce the following notations. For a circular order $\Cc$, let $-\Cc$ denote its reverse
circular order, \ie  for $\Cc=(n_1,n_2,\cdots,n_N)$,
we have $-\Cc=(n_N,n_{N-1},\cdots,n_1)$.
For a channel~$i$, let $i_{\Cc}^+$ denote the next channel in the circular order $\Cc$. For example,
for $\Cc=(1,2,\cdots,N)$, we have $i_{\Cc}^+=i+1$ for $1\le i<N$ and $N_{\Cc}^+=1$.

We present below the structure of the myopic policy. We assume first that the initial belief value
$\omega_i(1)$ of each channel is bounded between $p_{01}$ and $p_{11}$. In Appendix~B,
we show that when this condition on the initial belief values is violated, the same
structure holds for $t>2$. The only difference is that special care needs to be given to the second
slot. This can be seen from the belief update given in \eqref{eq:omega}.
Specifically, for any initial belief value, the updated belief of
each channel (observed or unobserved) in slot $t\ge 2$ is bounded between $p_{01}$ and $p_{11}$; a belief
value outside the interval of $[\min\{p_{01},p_{11}\},\max\{p_{01},p_{11}\}]$ can only occur in the first
slot as a given initial state, thus referred to as a transient belief state.

\vspace{0.5em}

\begin{theorem} {\it Structure of Myopic Policy.}\\
Let $\Omega(1)=[\omega_1(1),\cdots,\omega_N(1)]$ denote the initial belief vector.
Assume that $\omega_i(1)\in [\min\{p_{01},p_{11}\},\max\{p_{01},p_{11}\}]$ for all $i=1,2,\cdots,N$.
The circular channel order $\Cc(1)$ in slot~$1$ is
determined by a descending order of $\Omega(1)$ (\ie $\Cc(1)=(n_1,n_2,\cdots,n_N)$ implies that
$\omega_{n_1}(1)\ge\omega_{n_2}(1)\ge\cdots\ge\omega_{n_N}(1)$).
Let $\hat{a}(1)=\arg\max_{i=1,\cdots,N} \omega_i(1)$. The myopic action $\hat{a}(t)$ in slot $t$ ($t>1$) is given as follows.
\begin{itemize}
\item Case 1: $p_{11}\ge p_{01}$ and $\epsilon<\frac{p_{10}p_{01}}{p_{11}p_{00}}$
\end{itemize}
\begin{equation}
\hat{a}(t)=\left\{\begin{array}{ll}
\hat{a}(t-1), & \mbox{if } K_{\hat{a}(t-1)}(t-1)=1\\
\hat{a}(t-1)_{\Cc(t)}^+, & \mbox{if } K_{\hat{a}(t-1)}(t-1)=0 \\
\end{array}
\right.,
\label{eq:structure1}
\end{equation}
where $\Cc(t)=\Cc(1)$.
\begin{itemize}
\item Case 2: $p_{11}< p_{01}$ and $\epsilon<\frac{p_{00}p_{11}}{p_{01}p_{10}}$
\end{itemize}
\begin{equation}
\hat{a}(t)=\left\{\begin{array}{ll}
\hat{a}(t-1) & \mbox{if } K_{\hat{a}(t-1)}(t-1)=0\\
\hat{a}(t-1)_{\Cc(t)}^+ & \mbox{if } K_{\hat{a}(t-1)}(t-1)=1 \\
\end{array}
\right.,
\label{eq:structure2}
\end{equation}
where $\Cc(t)=\Cc(1)$ when $t$ is odd and $\Cc(t)=-\Cc(1)$ when $t$ is even.
\label{thm:structure}
\end{theorem}

\begin{proof}
See Appendix~A.
\end{proof}

\vspace{0.5em}

Theorem~\ref{thm:structure} along with Appendix~B shows that the basic structure of the myopic policy
is a round-robin scheme based on a circular ordering of the channels.
For $p_{11}\ge p_{01}$ (which corresponds to a positive correlation
between the channel states in two consecutive slots), the circular order is constant: $\Cc(t)=\Cc(1)$
in every slot~$t$, where $\Cc(1)$ is determined by a descending
order of the initial belief values. The myopic action is to stay in the same channel after
an ACK and switch to the next channel in the circular order after a NAK, provided that
the false alarm probability $\epsilon$ of the channel state detector is below a certain
value.

For $p_{11}< p_{01}$ (which corresponds to a negative correlation
between the channel states in two consecutive slots), the circular order is reversed in every slot:
$\Cc(t)=\Cc(1)$ when $t$ is odd and $\Cc(t)=-\Cc(1)$ when $t$ is even, where
the initial order $\Cc(1)$ is determined by the initial belief values. The myopic policy stays in the same channel
after a NAK; otherwise, it switches to the next channel in the {\it current} circular order $\Cc(t)$, which is
either $\Cc(1)$ or $-\Cc(1)$ depending on whether the current time $t$ is odd or even\footnote{An alternative way to see
the channel switching structure of the myopic policy is through the last visit to each channel (once every channel
has been visited at least once). Specifically, for $p_{11}\ge p_{01}$, when
a channel switch is needed, the policy
selects the channel visited the longest time ago. For $p_{11}< p_{01}$,
when a channel switch is
needed, the policy selects, among those channels to which the last
visit occurred an even number of slots ago, the one most recently
visited. If there are no such channels, the user chooses the channel
visited the longest time ago.}.

This simple structure suggests that the myopic sensing policy is particularly
attractive in implementation. Besides its simplicity, the myopic policy obviates
the need for knowing the channel transition probabilities and automatically tracks
variations in the channel model. %Note that when $p_{11}=p_{01}$, channel
%states become independent in time; all channel selections lead to the same
%performance. We thus expect that myopic sensing is robust to estimation errors
%in the order of $p_{11}$ and $p_{01}$, which usually occur when
%$p_{11}\approx p_{01}$. This has been confirmed by simulation results.

We point out that the structure of the myopic sensing policy in the presence of sensing errors
is similar to that under perfect sensing given in \cite{Zhao&Krish:07CogNet}. The proof, however,
is more involved since the observations here are acknowledgements and the state of the sensed channel
cannot be inferred with certainty from a NAK.

Theorem~\ref{thm:optimality} below shows that the myopic sensing policy with such a simple and
robust structure is, in fact, optimal for $N=2$.

\vspace{0.2em}

\begin{theorem} {\it Optimality of Myopic Policy.}\\
 For $N=2$, the myopic policy is optimal when $\epsilon<\frac{p_{10}p_{01}}{p_{11}p_{00}}$
for positively correlated channels ($p_{11}\ge p_{01}$) and $\epsilon<\frac{p_{00}p_{11}}{p_{01}p_{10}}$
for negatively correlated channels ($p_{11}< p_{01}$) when the initial belief values are bounded\footnote{Recall that
a belief value outside the interval of $[\min\{p_{01},p_{11}\},\max\{p_{01},p_{11}\}]$ is transient.
For any initial state, the belief values in slots $t\ge 2$ are bounded between $p_{01}$ and $p_{11}$.
As a consequence, Theorem~2 shows that when one or more of the initial belief values
are transient, the myopic
policy still provides the optimal actions in all slots except maybe the first slot.}
between $p_{01}$ and $p_{11}$.
\label{thm:optimality}
\end{theorem}

\vspace{0.5em}

\begin{proof}
See Appendix~B.
\end{proof}

\vspace{0.2em}

Numerical examples suggest that there exist similar conditions for
all $N$ under which the myopic policy is optimal. Proving this
conjecture turns out to be challenging. A recent
work~\cite{Javidi&etal:08ICC} has made progress towards proving a
corresponding conjecture under the assumption of perfect sensing, by
showing that the optimality holds for $N > 2$ under the condition
that $p_{11}>p_{01}$. Furthermore, it is shown
in~\cite{Javidi&etal:08ICC} that if the myopic policy is optimal
under the sum-reward criterion over a finite horizon, it is also
optimal for other criteria such as discounted and averaged rewards
over a finite or infinite horizon. These results may be extended to
the case with noisy observations, since the optimality proof given
in~\cite{Javidi&etal:08ICC} exploits the simple structure of the
myopic policy, which, as shown here, also holds with noisy
observations.

Both the structure and the optimality of the myopic policy require a certain level of reliability
of the channel state detector. When this level of reliability is not met, the simple structure of
the myopic policy may no longer hold, and the myopic actions need to be obtained from \eqref{eq:a*}
and the recursive belief update in \eqref{eq:omega}. The optimality of the myopic policy may also
be lost in this case. A more complex policy, for example, Whittle's index policy \cite{whittle},
may need to be sought after to achieve better performance.
This brings out an interesting tradeoff between the complexity of the detector at the physical layer
and the complexity of the sensing strategy at the Medium Access Control (MAC) layer. In particular, the reliability of a detector
(for example, an energy detector) can always be improved by increasing the sensing time so that a simple
and optimal policy---the myopic policy---can be employed. The caveat is the reduced transmission time
for a given slot length. Such a tradeoff can be complex and is beyond the scope of this technical note.

\vspace{-0.6em}
\section{Conclusion and Discussions}
\vspace{-0.5em}

We have established a simple structure of the myopic policy for channel selection in
an $N$-channel opportunistic communication system under an i.i.d. Gilbert-Elliot channel model.
The optimality of this simple myopic policy is proved for $N=2$ and conjectured for $N>2$. This is a non-trivial
extension of our previous results pertaining to the case of error-free
channel state detection~\cite{Zhao&Krish:07CogNet}, as noisy observations make it
challenging to maintain synchronous channel selection between the transmitter and its receiver.
This communication constraint adds an interesting twist to the resulting stochastic control
problem.

The optimality of the myopic policy in the context of opportunistic
communications may bear significance in the general context of
restless multi-armed bandit processes. While the classical bandit
problems can be solved optimally using the Gittins Index
\cite{gittins}, restless bandit problems are known to be PSPACE-hard
in general \cite{tsitsiklis}. Whittle proposed a Gittins-like
indexing heuristic for the restless bandit problems~\cite{whittle}
which is shown to be asymptotically optimal in certain limiting
regime \cite{weber}. Beyond this asymptotic result, relatively
little is known about the structure of the optimal policies for a
general restless bandit process. The optimality of the
myopic policy shown in this paper and \cite{Zhao&Krish:07CogNet}
suggests non-asymptotic conditions under which an index policy with
a semi-universal structure can
actually be optimal for restless bandit processes.

Approximation algorithms for restless bandit problems have also been
explored in the literature. In~\cite{Guha&Munagala:07FOCS}, Guha and Munagala
have developed a
constant-factor ($1/68$) approximation via LP relaxation for the same class of restless
bandit processes as considered in this paper. The difference is
that the model in~\cite{Guha&Munagala:07FOCS} allows for non-identical
channels but every channel is positively correlated. We point out that negatively
correlated processes are significantly harder to deal with due to the loss of
monotonicity in the belief updates (see \cite{Zhao&Krish:07CogNet}). In~\cite{Guha&etal:07},
Guha~\emph{et al.} have developed a factor~$2$ approximation policy for another
class of restless bandit problems (referred to as monotone bandits) via LP relaxation.
Raghunathan~\emph{et al.}~\cite{Raghunathan&etal:08INFOCOM} have also modeled multicast
scheduling in broadcast wireless LANs as a restless bandit problem
and provided a closed-form bound for the performance of Whittle's
index policy with respect to the optimal.

{\small
\section*{Appendix A: Proof of Theorem~\ref{thm:structure}}
\vspace{-0.5em}

We prove Theorem~\ref{thm:structure} by showing that the channel $\hat{a}(t)$ given by
\eqref{eq:structure1} and \eqref{eq:structure2} is indeed the channel with the largest
belief value in slot $t$. Specifically, we prove the following lemma.

\begin{lemma}
Let $\hat{a}(t)=i_1$ be the channel determined by \eqref{eq:structure1} for $p_{11}\ge p_{01}$
and by \eqref{eq:structure2} for $p_{11}< p_{01}$. Let $\Cc(t)=(i_1,i_2,\cdots,i_N)$ be the
circular order of channels in slot $t$, where we set the starting point to $\hat{a}(t)=i_1$.
We then have, for any $t\ge 1$,
\begin{equation}
\omega_{i_1}(t)\ge\omega_{i_2}(t)\ge\cdots\ge\omega_{i_N}(t),
\label{eq:A11}
\end{equation}
\label{lemma:structure}
\ie the channel given by \eqref{eq:structure1} and \eqref{eq:structure2} has the largest
belief value in every slot $t$.
\end{lemma}

To prove Lemma~\ref{lemma:structure}, we note the following properties of the operator $\Gamma(x)$ defined in \eqref{eq:omega}.
\begin{itemize}
\item[P1.] $\Gamma(x)$ is an increasing function for $p_{11}\ge p_{01}$
and a decreasing function for $p_{11}<p_{01}$.
\item[P2.] $\forall 0\le x\le 1$, $p_{01}\le \Gamma(x)\le p_{11}$ for $p_{11}\ge p_{01}$
and $p_{11}\le \Gamma(x)\le p_{01}$ for $p_{11}<p_{01}$.
\item[P3.] For $p_{11}\ge p_{01}$ and $\epsilon<\frac{p_{10}p_{01}}{p_{11}p_{00}}$, we have
$\Gamma(\frac{\epsilon\omega}{\epsilon\omega+(1-\omega)})\le \Gamma(\omega')$
$\forall p_{01}\le \omega,\omega'\le p_{11}$;
for $p_{11}< p_{01}$ and $\epsilon<\frac{p_{00}p_{11}}{p_{01}p_{10}}$, we have
$\Gamma(\frac{\epsilon\omega}{\epsilon\omega+(1-\omega)})\ge \Gamma(\omega')$
$\forall p_{11}\le \omega,\omega'\le p_{01}$.
\end{itemize}
P1 and P2 follow directly from the definition of $\Gamma(x)$. To show P3 for $p_{11}\ge p_{01}$,
it suffices to show $\frac{\epsilon\omega}{\epsilon\omega+(1-\omega)}\le p_{01}$ due to the
monotonically increasing property of $\Gamma(x)$ and the bound on $\omega'$. Noticing that
$\frac{\epsilon\omega}{\epsilon\omega+(1-\omega)}$ is an increasing function of both $\omega$
and $\epsilon$, we arrive at P3 by using the upper bounds on $\omega$ and $\epsilon$. Similarly,
we can show P3 for $p_{11}< p_{01}$.

We now prove Lemma~\ref{lemma:structure} by induction.
For $t=1$, \eqref{eq:A11} holds by the definition of $\Cc(1)$.
Assume that \eqref{eq:A11} is true for slot~$t$, where
$\Cc(t)=(i_1,i_2,\cdots,i_N)$ and $\hat{a}(t)=i_1$.
We show that it is also true for slot~$t+1$.

Consider first $p_{11}\ge p_{01}$. We have $\Cc(t+1)=\Cc(t)=(i_1,i_2,\cdots,i_N)$.
When $K_{i_1}(t)=1$, we have $\hat{a}(t+1)=\hat{a}(t)=i_1$
from \eqref{eq:structure1}. Since $\omega_{i_1}(t+1)=p_{11}$ achieves the upper bound of the belief values (see P2)
and the order of the belief values of the unobserved channels remains unchanged due to P1,
we arrive at \eqref{eq:A11} for $t+1$. When $K_{i_1}(t)=0$, we have $\hat{a}(t+1)=i_2$
from \eqref{eq:structure1}. We again have \eqref{eq:A11} by noticing that
$\omega_{i_1}(t+1)=\Gamma(\frac{\epsilon\omega_{i_1}(t)}{\epsilon\omega_{i_1}(t)+(1-\omega_{i_1}(t))})$ is
the smallest belief value in slot $t+1$ (see P3) and $\Cc(t+1)=(i_2,i_3,\cdots,i_N,i_1)$ when the starting point
is set to $\hat{a}(t+1)=i_2$.

For $p_{11}<p_{01}$, $\Cc(t+1)=-\Cc(t)=(i_1,i_N,i_{N-1},\cdots,i_2)$.
When $K_{i_1}(t)=0$, we have $\hat{a}(t+1)=\hat{a}(t)=i_1$
from \eqref{eq:structure2}. Since $\omega_{i_1}(t+1)=\Gamma(\frac{\epsilon\omega}{\epsilon\omega+(1-\omega)})$
is the largest belief value in slot $t+1$ (see P3)
and the order of the belief values of the unobserved channels is reversed due to P1, we have, from the induction assumption at~$t$,
\[
\omega_{i_1}(t+1)\ge\omega_{i_N}(t+1)\ge\omega_{i_{N-1}}(t+1)\ge\cdots\ge\omega_{i_2}(t+1),
\]
which agrees with \eqref{eq:A11} for~$t+1$ and $\Cc(t+1)=(i_1,i_N,i_{N-1},\cdots,i_2)$.
When $K_{i_1}(t)=1$, we have $\hat{a}(t+1)=i_N$
from \eqref{eq:structure2}. We again have \eqref{eq:A11} by noticing that $\omega_{i_1}(t+1)=p_{11}$ achieves
the lower bound of the belief values and $\Cc(t+1)=(i_N,i_{N-1},\cdots,i_2,i_1)$ when the starting point
is set to $\hat{a}(t+1)=i_N$. This concludes
the proof of Lemma~\ref{lemma:structure}, hence Theorem~\ref{thm:structure}.

\vspace{-0.6em}
\section*{Appendix B: Structure of the Myopic Policy under Transient Initial Belief States}
\vspace{-0.5em}

We now consider when one or more initial belief values are transient, \ie outside the interval
of $[\min\{p_{01},p_{11}\},\max\{p_{01},p_{11}\}]$.
Let $\Omega(1)=[\omega_1(1),\cdots,\omega_N(1)]$ denote the initial belief vector.
Without loss of generality, assume that $\omega_1(1)\ge\omega_2(1)\ge\cdots\ge\omega_N(1)$. Thus
$\hat{a}(1)=1$. Let $r$ denote the rank of $\frac{\epsilon\omega_1(1)}{\epsilon\omega_1(1)+(1-\omega_1(1))}$
in $\{\frac{\epsilon\omega_1(1)}{\epsilon\omega_1(1)+(1-\omega_1(1))}, \omega_2(1), \cdots,\omega_N(1)\}$
with $r=1$ when $\frac{\epsilon\omega_1(1)}{\epsilon\omega_1(1)+(1-\omega_1(1))}$ is the largest
and $r=N$ when it is the smallest.
When one or more of the initial belief values are transient, the myopic action $\hat{a}(t)$ in slot $t$ ($t>1$) is given as follows.
\begin{itemize}
\item Case 1: $p_{11}\ge p_{01}$ and $\epsilon<\frac{p_{10}p_{01}}{p_{11}p_{00}}$
\begin{itemize}
\item If $K_{\hat{a}(1)}(1)=1$, the myopic action $\hat{a}(t) ~(t>1)$ follows the same structure
given by \eqref{eq:structure1} with $\Cc(1)=(1,2,\cdots,N)$.
\item If $K_{\hat{a}(1)}(1)=0$, the myopic action in slot $t=2$ is $\hat{a}(2)=1$ when $r=1$
and $\hat{a}(2)=2$ when $r>1$. The myopic action $\hat{a}(t)$ for $t>2$ follows the same structure
given by \eqref{eq:structure1} with $\Cc(1)=(1,2,\cdots,N)$ when $r=1$ and $\Cc(1)=(2,3,\cdots,r, 1, r+1, r+2,\cdots, N)$
when $r>1$.
\end{itemize}
\item Case 2: $p_{11}< p_{01}$ and $\epsilon<\frac{p_{00}p_{11}}{p_{01}p_{10}}$
\begin{itemize}
\item If $K_{\hat{a}(1)}(1)=1$, the myopic action $\hat{a}(t) ~(t>1)$ follows the same structure
given by \eqref{eq:structure2} with $\Cc(1)=(1,2,\cdots,N)$.
\item If $K_{\hat{a}(1)}(1)=0$, the myopic action in slot $t=2$ is $\hat{a}(2)=1$ when $r=N$
and $\hat{a}(2)=N$ when $r<N$. The myopic action $\hat{a}(t)$ for $t>2$ follows the same structure
given by \eqref{eq:structure2} with $\Cc(1)=(1,2,\cdots,N)$ when $r=1$ and $\Cc(1)=(2,3,\cdots,r, 1, r+1, r+2,\cdots, N)$
when $r>1$.
\end{itemize}
\end{itemize}
The above modification can be easily proved based on P1 and P2 given in Appendix~A.

\vspace{-0.6em}
\section*{Appendix C: Proof of Theorem~\ref{thm:optimality}}
\vspace{-0.5em}

Let $\hat{V}_t(\Omega)$ denote the total expected reward obtained under the myopic policy
starting from slot $t$, and $\hat{V}_t(\Omega;a)$ the total expected reward obtained by action $a$
in slot $t$ followed by the myopic policy in future slots.
The proof is based on the following lemma which applies to a general POMDP.

\begin{lemma}
For a $T$-horizon POMDP, the myopic policy is optimal if for $t=1,\cdots,T$,
\begin{equation}
\hat{V}_t(\Omega)\ge \hat{V}_t(\Omega;a),~~~\forall a,\Omega.
\label{eq:C0}
\end{equation}
\label{lemma:pomdp}
\end{lemma}

\vspace{-1em}

Lemma~\ref{lemma:pomdp} can be proved by reverse induction, where the initial condition of the
optimality of the myopic action in that last slot $T$ is straightforward.

We now prove Theorem~\ref{thm:optimality}. Considering all channel state realizations in slot $t$,
we have
\begin{equation}
\hat{V}_t(\Omega; a)=(1-\epsilon)\omega_a+\sum_{s_1,s_2\in\{0,1\}} \Pr[\Sbf(t)=[s_1,s_2]~|~\Omega(t)]
\hat{V}_{t+1}(\Tc(\Omega(t)|a,s_a)~|~\Sbf(t)=[s_1,s_2]),
\label{eq:conditionalV}
\end{equation}
where $\hat{V}_{t+1}(\Tc(\Omega(t)|a,s_a)~|~\Sbf(t)=[s_1,s_2])$ is the conditional reward obtained starting
from slot $t+1$ given that the system state in slot $t$ is $[s_1,s_2]$.
Next, we establish two lemmas regarding the conditional value function of the myopic policy.

\vspace{0.2em}

\begin{lemma}
Under the conditions of Theorem~1, the expected total remaining reward
starting from slot $t$ under the myopic policy is determined by the action $a(t-1)$ and the system
state $\Sbf(t-1)$ in slot $t-1$, hence independent of the belief vector $\Omega(t)$ at the beginning of slot $t$, \ie
\[
\hat{V}_{t}(\Tc(\Omega(t-1)|a,s_a)~|~\Sbf(t-1)=[s_1,s_2])=\hat{V}_{t}(\Tc(\Omega'(t-1)|a,s_a)~|~\Sbf(t-1)=[s_1,s_2]).
\]
Adopting the simplified notation of $\hat{V}_t(a(t-1)|\Sbf(t-1)=[s_1,s_2])$, We further have
\begin{equation}
\hat{V}_t(a(t-1)=1|\Sbf(t-1)=[s_1,s_2])=\hat{V}_t(a(t-1)=2|\Sbf(t-1)=[s_2,s_1]).
\label{eq:prop3b}
\end{equation}
\end{lemma}

\vspace{0.2em}

\begin{proof} Given $a(t-1)$ and $\Sbf(t-1)$, the myopic actions in slots $t$ to $T$, governed by the structure given in
Theorem~\ref{thm:structure}, are fixed
for each sample path of system state and observation, independent of $\Omega(t)$. As a consequence, the total reward
obtained in slots $t$ to $T$ for each sample path is independent of $\Omega(t)$, so is the expected total reward.
\eqref{eq:prop3b} follows from the statistically identical assumption of channels.
\end{proof}

\vspace{0.2em}

\begin{lemma}
Under the conditions of Theorem~1, we have, $\forall t,a$,
\begin{equation}
\left|\hat{V}_t(a(t-1)=a|\Sbf(t-1)=[1,0])-\hat{V}_t(a(t-1)=a|\Sbf(t-1)=[0,1])\right|\le (1-\epsilon).
\label{eq:B1}
\end{equation}
\end{lemma}

\vspace{0.2em}

\begin{proof} Based on \eqref{eq:prop3b},
it suffices to consider $a(t-1)=1$.
We prove for $p_{11}<p_{01}$ by reverse induction. The proof for $p_{11}>p_{01}$ is similar.
The inequality in \eqref{eq:B1} holds for $t=T$ since $(1-\epsilon)$ is the maximum expected reward
that can be obtained in one slot.
Assume that the inequality holds for $t+1$. We show that it holds for $t$.
Consider first $\hat{V}_t(a(t-1)=1|\Sbf(t-1)=[1,0])$. With probability $1-\epsilon$, the user successfully
identifies that channel $1$ is in the good state in slot $t-1$ and receives an acknowledgement at the end
of slot $t-1$. According to the structure of the myopic policy, the user switches channel in slot $t$, \ie
$a(t)=2$. The expected immediately reward in slot $t$ is thus $p_{01}(1-\epsilon)$ since the state of
channel $2$ in slot $t-1$ is $0$. We thus arrive at the first term of~\eqref{eq:Vt1}, where
$\hat{V}_t(a(t-1)=1|\Sbf(t-1)=[1,0])$ is given by the summation
of $p_{01}(1-\epsilon)$ and the future reward starting from slot $t+1$ conditioned on all four possible
system states in slot $t$. With probability $\epsilon$, a false alarm occurs in slot $t-1$, resulting in
a NAK. The user thus stays in channel $1$ in slot $t$: $a(t)=1$. We thus arrive at the second term of~\eqref{eq:Vt1}.
Similarly, we obtain $\hat{V}_t(a(t-1)=1|\Sbf(t-1)=[0,1])$ as given in~\eqref{eq:Vt2}, which follows from the
fact that a NAK occurs in slot $t-1$ due to the given bad state of the chosen channel $1$.
\begin{eqnarray}
\hat{V}_t(1|[1,0])&=&(1-\epsilon)\left\{p_{01}(1-\epsilon)+p_{10}p_{00}\hat{V}_{t+1}(2|[0,0])
+p_{11}p_{01}\hat{V}_{t+1}(2|[1,1])%\right. \nn\\
%& & \hspace{0.6in}\left.
+p_{11}p_{00}\hat{V}_{t+1}(2|[1,0])+p_{10}p_{01}\hat{V}_{t+1}(2|[0,1])\right\}\nn\\
&& +\epsilon\left\{p_{11}(1-\epsilon)+p_{10}p_{00}\hat{V}_{t+1}(1|[0,0])+p_{11}p_{01}\hat{V}_{t+1}(1|[1,1])%\right. \nn\\
%&& \hspace{0.3in} \left.
+p_{11}p_{00}\hat{V}_{t+1}(1|[1,0])+p_{10}p_{01}\hat{V}_{t+1}(1|[0,1])\right\}\label{eq:Vt1}\\[0.5em]
\hat{V}_t(1|[0,1])&=&p_{01}(1-\epsilon)+p_{00}p_{10}\hat{V}_{t+1}(1|[0,0])+p_{01}p_{11}\hat{V}_{t+1}(1|[1,1])%\nn\\
%& &
+p_{11}p_{00}\hat{V}_{t+1}(1|[0,1])+p_{10}p_{01}\hat{V}_{t+1}(1|[1,0])\label{eq:Vt2}
\end{eqnarray}
Applying \eqref{eq:prop3b} and the upper bound on $\epsilon$, we have
\begin{eqnarray}
&&\left|\hat{V}_t(1|[0,1])-\hat{V}_t(1|[1,0])\right|\nn\\
&\le&(1-\epsilon)p_{01}-(1-\epsilon)(\epsilon p_{11}+(1-\epsilon)p_{01})
+\epsilon \left|\hat{V}_{t+1}(1|[1,0])-\hat{V}_{t+1}(1|[0,1]\right|(p_{10}p_{01}-p_{11}p_{00})\nn\\
&\le&2(1-\epsilon) \epsilon (p_{01}-p_{11})\nn\\
&\le&2(1-\epsilon) \frac{p_{00}p_{11}}{p_{01}p_{10}}(p_{01}-p_{11})\nn\\
&<&(1-\epsilon),\nn
\end{eqnarray}
where the last inequality follows from  $(p_{01}-p_{11})\frac{p_{11}}{p_{01}}\le\frac{1}{4}$
and $\frac{p_{00}}{p_{10}}<1$.
\end{proof}

\vspace{0.2em}

We now show that \eqref{eq:C0} in Lemma~\ref{lemma:pomdp} holds.
Consider $\Omega(t)=[\omega_1(t),\omega_2(t)]$ with $\omega_1(t)>\omega_2(t)$, \ie the myopic action in slot $t$ is $a(t)=1$.
Applying \eqref{eq:prop3b} and Lemma~4 to \eqref{eq:conditionalV}, we have
\begin{equation}
\hat{V}_t(\Omega; a=1)-\hat{V}_t(\Omega;a=2)
=(\omega_1-\omega_2)(1-\epsilon+\hat{V}_{t+1}(1|[1,0])-\hat{V}_{t+1}(1|[0,1]))
\ge 0.\nn
\end{equation}
}

\vspace{-2em}
%%%%%%%%%% References %%%%%%%%%%%%%%%%%%%%%%%%%%%%%%%%%%%%%%%%%%%%%%%%%%
\bibliographystyle{ieeetr}
{\small
%\bibliography{\HOME/Reference/Bibs/Journal,%
%\HOME/Reference/Bibs/Conf,%
%\HOME/Reference/Bibs/Misc,%
%\HOME/Reference/Bibs/Book,\HOME/Reference/Bibs/CAREER}

}

\end{document}